\documentclass[a4paper]{article}

 \usepackage{graphicx}

\usepackage[tight]{subfigure}

\usepackage{amssymb}

\makeatletter
\renewcommand\thefigure{\@arabic\c@figure}

\long\def\@makecaption#1#2{%
\vskip\abovecaptionskip
\sbox\@tempboxa{#1.\hskip 1em#2}%
\ifdim \wd\@tempboxa >\hsize
  #1.\hskip 1em#2\par
\else
  \global \@minipagefalse
  \hb@xt@\hsize{\hfil\box\@tempboxa\hfil}%
\fi \vskip\belowcaptionskip} \makeatother

\begin{document}

\title{Evolution of Cooperation among Mobile Agents}
\author{Zhuo Chen$^{1*}$, Jianxi Gao$^1$, Yunze Cai$^2$ and Xiaoming Xu$^{1,2,3}$\\
$^1$Shanghai Jiao Tong University, Shanghai, China\\
$^2$University of Shanghai For Science and Technology, Shanghai, China\\
$^3$Shanghai Academy of Systems Science, Shanghai, China\\
$^*$jeffchen\_ch@yahoo.com.cn}
\maketitle

\begin{abstract}

We study the effects of mobility on the evolution of cooperation among mobile players, which imitate collective motion of biological flocks and interact with neighbors within a prescribed radius $R$. Adopting the prisoner's dilemma game and the snowdrift game as metaphors, we find that cooperation can be maintained and even enhanced for low velocities and small payoff parameters, when compared with the case that all agents do not move.
But such enhancement of cooperation is largely determined by the value of $R$, and for modest values of $R$, there is an optimal value of velocity to induce the maximum cooperation level. Besides, we find that intermediate values of $R$ or initial population densities are most favorable for cooperation, when the velocity is fixed. Depending on the payoff parameters, the system can reach an absorbing state of cooperation when the snowdrift game is played. Our findings may help understanding the relations between individual mobility and cooperative behavior in social systems.
\end{abstract}

\section{Introduction}
Cooperation is commonly observed throughout biological systems, animal kingdoms and human societies.
But from a Darwinian viewpoint, cooperators are at a disadvantage in natural selection,
because they increase the fitness of others at the cost of their own survival and reproduction \cite{Okasha}.
In a broad range of disciplines, understanding the emergence of cooperation is a fundamental problem,
which is often studied within the framework of evolutionary game theory.

The prisoner's dilemma (PD) game and the snowdrift game are commonly used two person games with two strategies,
cooperation (C) and defection (D).
Mutual cooperation pays each a reward $R$, while mutual defection brings each a punishment $P$.
When one defector meets one cooperator, the former gains the temptation $T$
while the latter obtains the sucker's payoff $S$.
The PD is defined by the payoffs, if $T>R>P>S$ and $2R>S+T$.
In a single round of the PD, though the individual interest can be maximized by defection, the collective payoff achieves the maximum only when both players cooperate.
Hence the dilemma arises.
As an alternative model to study cooperative behavior, the SD is produced when $T>R>S>P$.
In contrast with the PD, the best strategy of the SD depends on the co-player: to defect if the opponent cooperates, but to cooperate if the opponent defects. Under replicator dynamics in well-mixed populations, defection is the only evolutionarily stable strategy in the PD, while cooperators may coexist with defectors in the SD. Note in the SD, the average population payoff at evolutionary equilibrium is smaller than that when everyone plays C \cite{Doebeli2005}. Thus SD is still a social dilemma.

One of possible mechanisms accounting for the establishment of cooperation is the so-called network reciprocity \cite{Nowak2006}. Discarding the well-mixed assumption for populations, this theory focuses on how spatial structure affects the evolution of cooperation.
Axelrod first suggested to locate individuals on the two-dimensional array, where interactions only happened within local neighborhoods. Nowak and May developed this idea later, showing that unconditional cooperators could survive by forming clusters \cite{Nowak1993}. These pioneering studies have triggered an intensive investigation of spatial games, yielding enormous combinations of evolutionary rules, graphs and game models.
In Ref. \cite{Szabo1998},  the effect of noise is incorporated in the strategy adoption, and Darwinian selection of the noise level favors a specific parameter value that induces the highest level of cooperation \cite{Szolnoki2009a,Szabo2009}.
Diversity is another role facilitating cooperation, which takes various forms as heterogeneous graphs \cite{Santos2008}, preferential imitations \cite{Yang2009}, reproduction probabilities \cite{Wu2009}, individual rationality \cite{Chen2009}, fitness \cite{Perc2008} or behavioral preferences \cite{VanSegbroeck2009}. Since connectivity structures in the real world  are far more than regular lattices, there are many interests in the impact of complex topologies on cooperative behavior \cite{Santos2005,Gomez-Gardenes2007,Ren2007,Chen2008,Du2009,Pena2009}.
The co-evolution of strategies and individual traits, such as teaching activities \cite{Szolnoki2008,Szolnoki2009}, learning rules \cite{Moyano2009,Cardillo2010} and social ties \cite{Zimmermann2004,Santos2006,Szolnoki2008a,Chen2009a,Fu2008}, constitutes a key mechanism for the sustainability of cooperation.
Interestingly, cooperators can benefit from the continuous supply of new players \cite{Poncela2008,Poncela2009}, and the strategy-independent evolution of networks can evoke powerful mechanisms to promote cooperation \cite{Szolnoki2009c,Szolnoki2009b}.
More details about spatial evolutionary games can be found in Ref. \cite{Doebeli2005,Nowak2006,Szabo2007,Perc2010} and references therein.

Mobility of individuals is responsible for various spatiotemporal dynamics on geographical scales,
such as the spread of infectious diseases and wireless viruses \cite{Kleinberg2007}.
And statistical properties of human motion have attracted much interest in recent years \cite{Brockmann2006,Gonzalez2008,Song2010}.
Indeed, the motion of individuals is an important characteristic of social networks \cite{Gonzalez2006}.
Though it is often neglected, the effects of mobility on the evolution of cooperation vary with movement forms and population structures.
Vainstein et al. \cite{Vainstein2007} considered a random diffusive process in a population of agents with pure strategies,
 where each agent can jump to a nearest empty site with a certain probability.
It was found that cooperation can be enhanced by the movement of players,
provided that the mobility parameter is kept with a certain range.
The weak form of the PD adopted in Ref. \cite{Vainstein2007} was later extended to other games \cite{Guan2007,Sicardi2009},and it was found that cooperation in the SD is not so often inhibited as that reported in Ref. \cite{Hauert2004}.
Besides, the movement of players may take an adaptive form for payoffs or neighbors, and contingent mobility is often expected to enhance cooperation. Aktipis \cite{Aktipis2004} proposed a walk-away strategy to avoid repeated interactions with defectors, which outperforms complex strategies under a number of conditions. Helbing and Yu introduced the success-driven migration, in which players determine destinations through fictitious play \cite{Helbing2009}.
Besides, individuals can decide when to move based on the number of neighboring defectors \cite{Jiang2010}.

The synchronised motion of animal groups, such as fish schools and bird flocks, is an intriguing phenomenon, which can be modeled by systems of self-driven agents \cite{Vicsek1995,Nagy2007,Dossetti2009}.
Recently, the model by Vicsek et al. has gained much attention for minimalism styles and rich dynamics \cite{Vicsek1995}.
Here we combine the Vicsek model with evolutionary games,
focusing on the effect of mobility on the evolution of cooperation.
We reserve well-known elements like direction alignment and circular neighborhoods, ignoring the influence of angular noise on the update of velocity.
We also cancel the periodic boundary conditions for simplicity,
which can strongly affect the system behavior at the large velocity regime \cite{Nagy2007}.
Thus when players move,
the system is split into some disconnected groups, within which agents move toward the same direction.
Note in some social systems, individuals do divide into groups according to race, wealth, age, and so on.
We think that the aggregation of individuals partly reflects the community structure in social networks.
In Ref. \cite{Chen2011}, we have investigated an evolutionary PD game in a Vicsek-like model, where each agent plays with constant number of neighbors. We have found that cooperation can be maintained and even enhanced by the motion of players, provided that certain conditions are fulfilled. In the current work, we will check the robustness of our conclusions, when each agent plays the PD game with those individuals within a certain distance. Besides, we will study how mobility affects the outcome of the SD game.

\section{The Model}
We consider a system with $N$ autonomous agents, which have positions $x_i(t)$ and move synchronously with velocities $\overrightarrow{V_i}(t)$ in a two-dimensional plane.
The velocity $\overrightarrow{V_i}(t)$ of the agent $i$ is
characterized by a fixed absolute velocity $v$ and an angle $\theta_i(t)$ indicating the direction of motion.
When $t=0$,
all agents are randomly distributed in an $L\times L$ square without boundary restrictions.
Rather than fixed within a periodic domain, individuals can cross the border of the square when $t>0$, and move in the whole plane. The square only represents the initial distribution of individuals with a density $\rho=N/L^2$.
Besides, initial moving directions of agents, $\theta_i(0)$, are uniformly distributed in the interval $[0,2\pi)$.
At each time step, the $i$th agent updates its position according to
\begin{equation}
  x_i(t+1)=x_i(t)+\overrightarrow{V_i}(t)\Delta t.
\end{equation}
Here $\Delta t$ is set to 1 between two updates on the positions.

To simulate the process of direction
alignment in flocks, the angle $\theta_i(t)$ of the agent $i$ is updated
according to the average direction of nearby neighbors \cite{Vicsek1995}. Then we have
\begin{equation}
\theta_i(t+1)=arctan\frac{sin\theta_i(t)+\sum_{j\in
W_i(t)}sin\theta_j(t)}{cos\theta_i(t)+\sum_{j\in
W_i(t)}cos\theta_j(t)},
\end{equation}
where $W_i(t)$ denotes the neighbors set of the agent $i$ at time $t$.
Here $W_i(t)$ is defined
as agents in the spherical neighborhood of the radius $R$ centered on the agent $i$,
\begin{equation}
  W_i(t)=\{j|\mid x_j-x_i\mid<R,j\in N,j\neq i\},
\end{equation}
where $|\bullet|$ denotes the Euclidean distance between j and i in two-dimensional
space. And we assume that each agent has the same radius.

The equations given above characterize the motion of agents.
When moving in the plane, the agents also play games in pairs.
For the PD, we take a re-scaled form suggested by Nowak et al. \cite{Nowak1993} as
\begin{equation}
  \mathbf{A}=\left(
  \begin{array}{ccc}
    1 & 0 \\
    b & 0
  \end{array}
  \right),
\end{equation}
where $b$ denotes the temptation to defect, and $1<b<2$.
And for the SD, we take a simplified form as
\begin{equation}
  \mathbf{A}=\left(
  \begin{array}{ccc}
    1 & 1-r \\
    1+r & 0
  \end{array}
  \right),
\end{equation}
where $r$ denotes the cost-to-benefit ratio of mutual cooperation, and $0<r<1$.
The strategy $s_i$ of the agent $i$, cooperation or defection, can be denoted by a unit vector
$(1,0)^T$ or $(0,1)^T$ respectively.
During the evolution of strategies,
an normalized payoff is calculated to exclude the effect coming from different degrees of players,
\begin{equation}
  P_i=\frac{\sum_{j\in W_i(t)}s_i^TAs_j}{\mid W_i(t) \mid},
\end{equation}
where  $|\bullet |$  represents the size of $W_i(t)$, and $A$ is the payoff matrix.
Afterward, every agent compares its income with that of its neighbors,
following the strategy which owns the highest payoff among its neighbors and itself \cite{Nowak1993}.

The system begins with an equal percentage of cooperators and defectors.
At each step, all agents collect payoffs and update strategies, and next, they modify positions and directions.
The time scale that characterizes the evolution of strategies is the same
as  the time scale that represents the motion of players.
This process is repeated until the system reaches equilibrium.

\begin{figure}[htbp]
\centering
\subfigure[]{\label{overview}
\includegraphics[height=4.6cm,width=5cm]{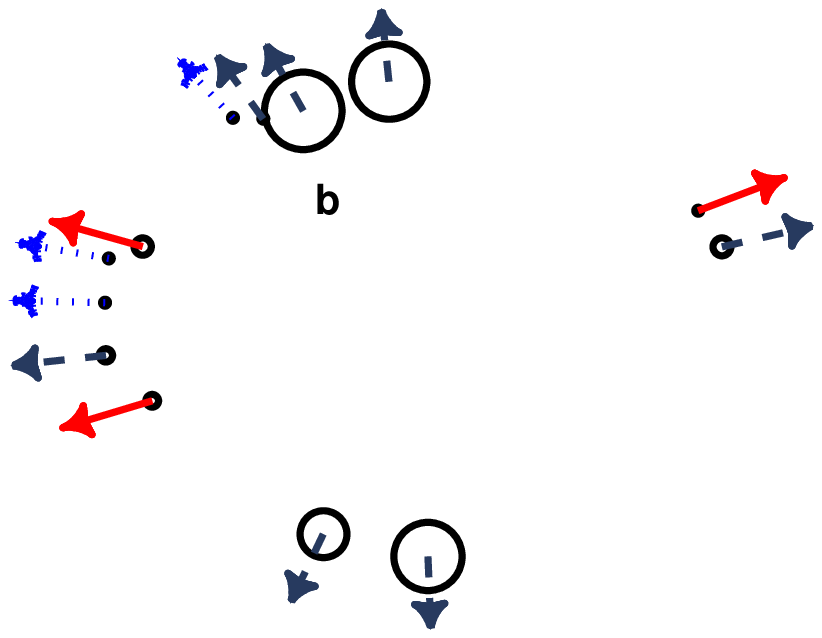}}
\subfigure[]{\label{detail}
\includegraphics[height=4.6cm,width=5cm]{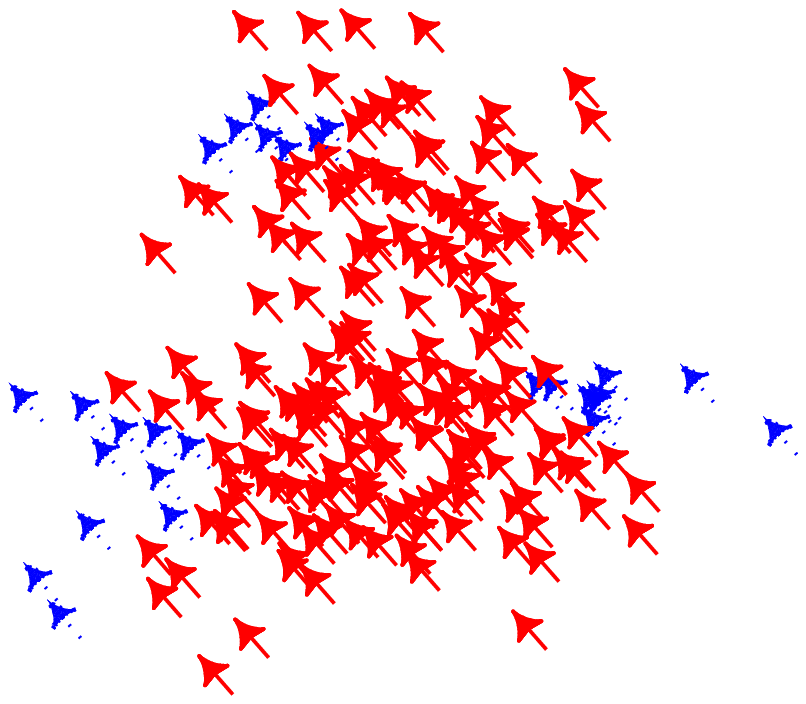}}
\subfigure[]{\label{typicalevo}
\includegraphics[height=5cm,width=5cm]{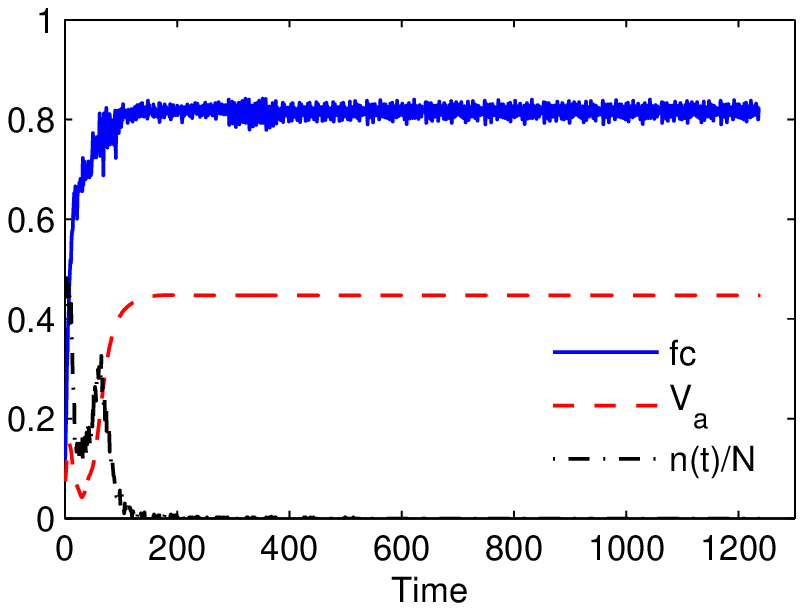}}
\caption{
(a)Snapshots of equilibrium configurations in the PD ($b=1.25$, $R=0.5$ and $v=0.05$).The trajectory of the center of each group is denoted by an arrow, which also indicates the group composition (solid arrow, all C; dotted arrow, all D; dashed arrow, coexistence of C and D). And the area of each circle is proportional to the group size. (b)Magnification of the labeled group (solid arrow, cooperator; dotted arrow, defector). (c)Representative time evolutions of $fc$, $n(t)/N$ and $V_a$ for the PD ($b=1.25$, $R=0.5$ and $v=0.05$). The data are obtained in one realization.}
\end{figure}

Fig. \ref{overview} illustrates the segregation of players at equilibrium. And players in the same group can move coherently, as shown in Fig. \ref{detail}.
During the process of direction alignment, the movement of players may lead to time-variant neighborhoods.
The total number of new neighbors appearing at time $t$ can be calculated as
\begin{equation}
  n(t)=\sum_{i=1}^{N}|W_i(t)-W_i(t)\bigcap W_i(t-1)|,
\end{equation}
where $|\bullet |$  represents the set size.
Fig. \ref{typicalevo} shows typical evolution of $n(t)$, which is divided by $N$ for normalization.
Besides, we also plot the evolution of the cooperator frequency $fc$ and the average normalized velocity $V_a$ \cite{Vicsek1995} for comparison.
When $t>500$, one can see that $n(t)$ decreases to $0$, and $V_a$ reaches a steady value.
These findings imply that a static interaction network has been constructed with fixed neighborhoods and velocities of players. When $t>1000$, $fc$ fluctuates stably. Then, the equilibrium frequency of cooperators can be obtained by averaging over a long period.

Simulations are carried out in a system with $N=1000$, $L=10$.
To ensure fixed topology of the interaction network, the evolution of $n(t)$ is monitored after a suitable relaxation time, which is varied from $5000$ to $10^5$ time steps and dependent on the values of $R$, $v$ and $L$.
If $n(t)\leq 1$, and this condition can hold for $q=1000$ time steps,
the network will be treated as a static one.
Then the equilibrium frequency of cooperators is evaluated by averaging over the last 1000 generations.
All data points shown in each figure are acquired by averaging over 200 realizations of independent initial states.
\section{Results and Discussions}
\begin{figure}[htbp]
\centering
\subfigure[]{ \label{fc-b:subfig:a}
\includegraphics[height=5cm,width=5cm]{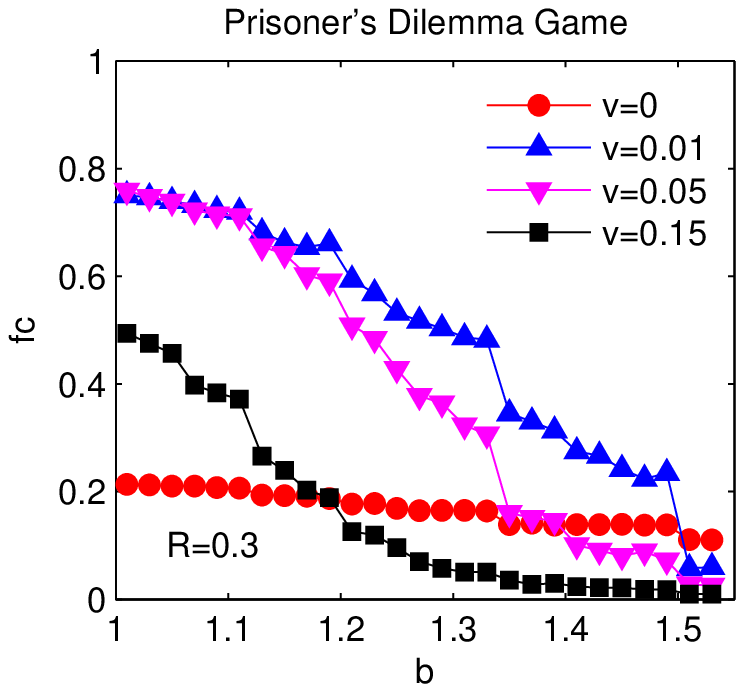}}
\subfigure[]{\label{fc-b:subfig:c}
\includegraphics[height=5cm,width=5cm]{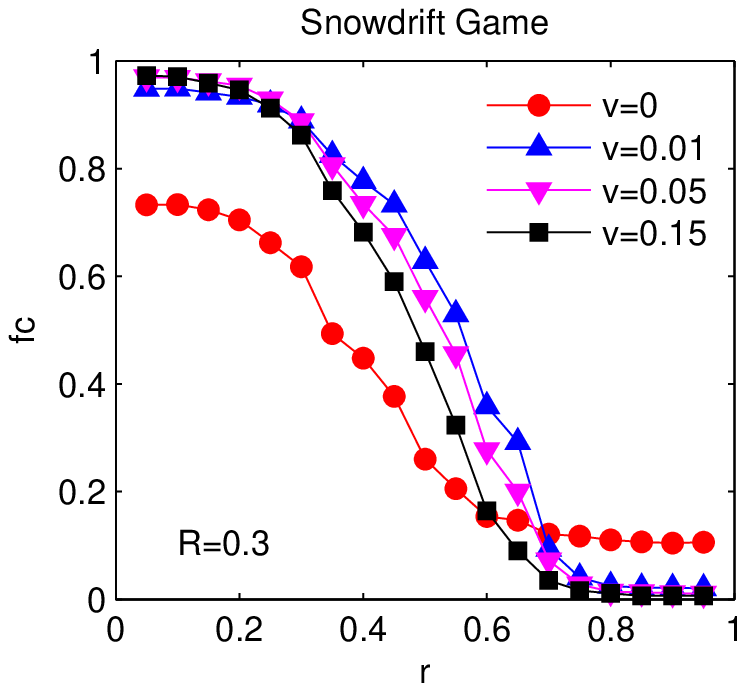}}
\caption{The cooperator frequency $fc$ versus the payoff parameters $b$ (PD) and $r$ (SD)
 for different values of $v$.
} \label{fc-b:subfig}
\end{figure}

Fig. \ref{fc-b:subfig} shows the fraction of cooperators $fc$ as a function of the payoff parameter,
$b$ for the PD and $r$ for the SD,
under different values of $v$ when $R$ is fixed.
Clearly, the cooperation level decreases with $b$ and $r$ in both games, no matter what $v$ is.
Compared with the static case ($v=0$), it is worth noting that cooperation is greatly enhanced in a large region of $b$ ($r$), when players are allowed to move with a low velocity (for example, $v=0.01$).
As shown in Fig. \ref{fc-b:subfig},
the proportion of cooperators for $v>0$ is higher than that for $v=0$,when $b<1.17$ or $r<0.6$,
and $fc$ can even approach $1$ in the SD.
But such an enhancement of cooperation can only be observed for small values of $b$ ($r$),
as the velocity increases from $0.01$ to $0.15$.
Indeed, a rapid decrease of the cooperator frequency can be observed in both games when $v=0.15$.

\begin{figure}[htbp]
\centering
\subfigure[]{\label{fcv:a}
\includegraphics[height=4.5cm,width=4.3cm]{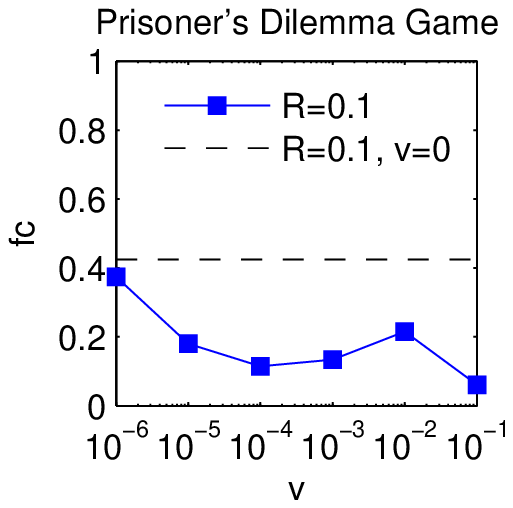}}
\subfigure[]{\label{fcv:b}
\includegraphics[height=4.5cm,width=4.3cm]{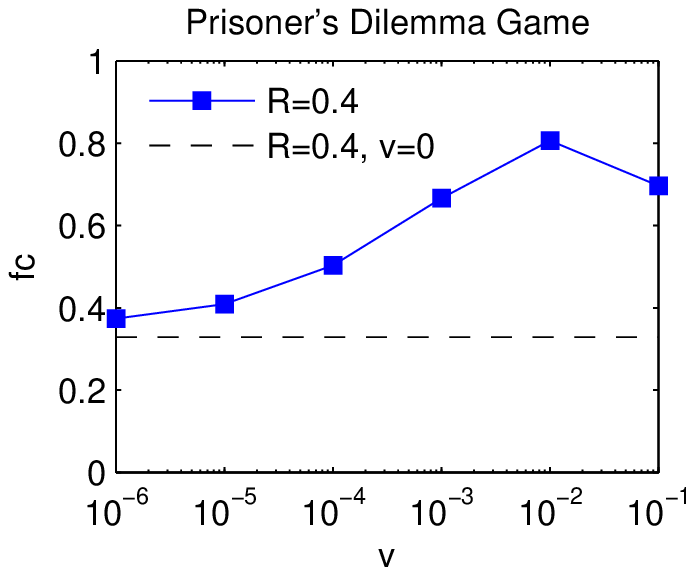}}
\subfigure[]{\label{fcv:c}
\includegraphics[height=4.5cm,width=4.3cm]{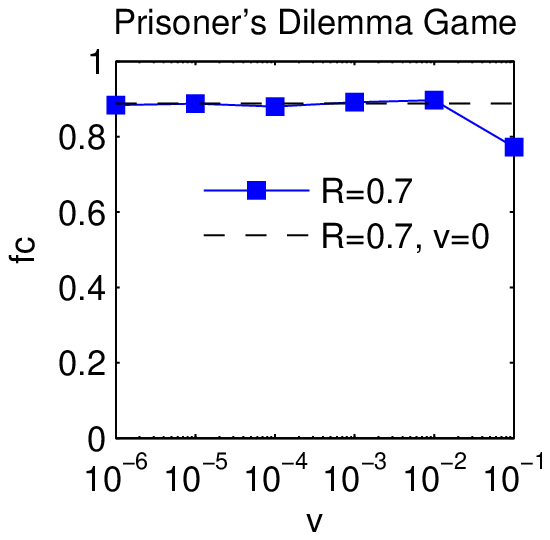}}
\subfigure[]{\label{fcv:d}
\includegraphics[height=4.5cm,width=4.3cm]{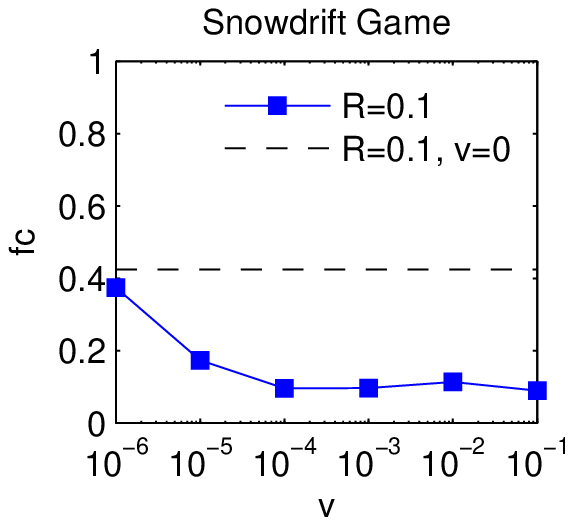}}
\subfigure[]{\label{fcv:e}
\includegraphics[height=4.5cm,width=4.3cm]{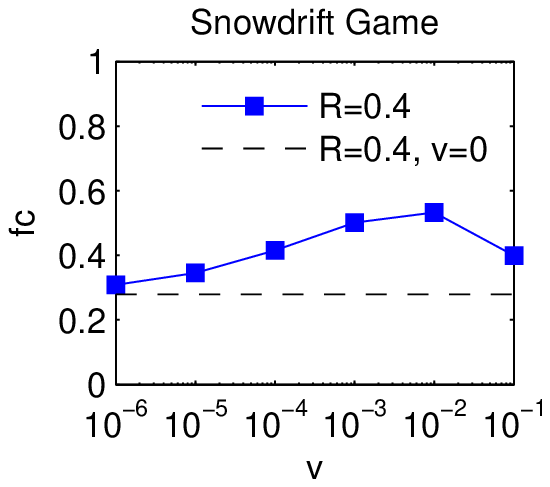}}
\subfigure[]{\label{fcv:f}
\includegraphics[height=4.5cm,width=4.3cm]{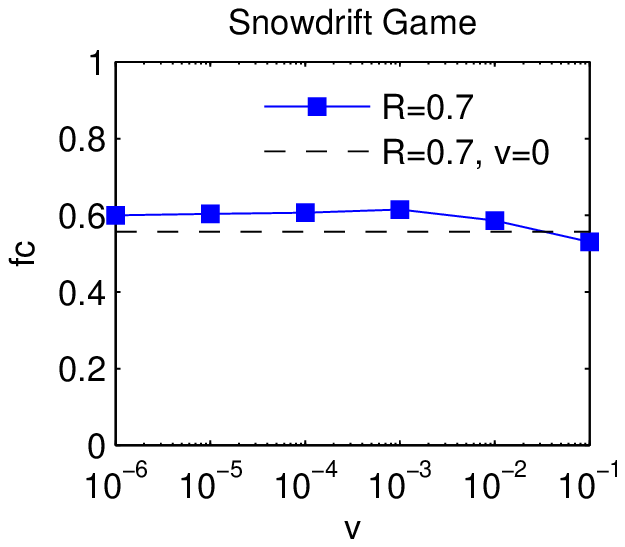}}
\caption{The frequency of cooperators $fc$ is plotted against the absolute velocity $v$ for different values of $R$, while the corresponding value of $fc$ for $v=0$ is presented by a dashed horizontal line in each figure. The temptation $b$ is set to $1.15$ (PD), and the cost-to-benefit ratio $r$ is set to $0.6$ (SD). A logarithmic scale is used for the $X$ axis.}\label{fcv}
\end{figure}

To clarify the effects of $v$ on the cooperator frequency, Fig. \ref{fcv} presents the dependence of the cooperation level $fc$ on the absolute velocity $v$ for different values of $R$.
It displays that the fraction of cooperators for $v=10^{-6}$ is very close to that for $v=0$,
irrespective of the value of $R$.
Meanwhile, it can be found that whether the movement of players promotes cooperation is largely determined by the value of $R$.
As shown in Fig. \ref{fcv:a} and Fig. \ref{fcv:d}, the movement of players fails to promote cooperation for small $R$.
One can see that the maximum of $fc$ for $R=0.1$ appears at $v=10^{-6}$ in both games,
and the cooperation level is lower than that for $v=0$ over the entire range of $v$.
For large $R$, the enhancement of cooperation resulting from the movement of players is quite limited or even disappeared.
As illustrated in Fig. \ref{fcv:c} and Fig. \ref{fcv:f}, when $v\leq 10^{-2}$, the curve of $fc$ nearly coincides with the result for $v=0$ in the PD, and only a tiny increase of $fc$ can be observed in the SD.
When $R=0.4$, however, the introduction of mobility can significantly improve cooperation in both games.
As shown in Fig. \ref{fcv:b} and Fig. \ref{fcv:e}, the cooperator frequency for $v>0$ is higher than that for $v=0$ in the whole region of $v$, and there is a maximum of $fc$ at $v=10^{-2}$.
The resonance-like phenomenon also implies that for a fixed $b$ ($r$), decreasing the value of $v$ cannot always promote cooperation.

Before moving forward, we would like to add some remarks about the above result.
Previous work has shown that cooperation is not only possible but may even be enhanced by the non-contingent movement of players when compared with the static case \cite{Vainstein2007,Dai2007,Meloni2009,YangArx}.
And our result provides another example that helps understand the universal role of mobility in the evolution of cooperation.
Particularly, Meloni and his colleagues \cite{Meloni2009} investigated how cooperation emerges in a population of PD players, which move in a square plane with periodic boundary conditions. The final outcome of the system has only two possibilities, all-C or all-D, when the neighborhood of each player is defined by a fixed radius $R$. The authors claimed that the movement of players can promote cooperation when the temptation to defect and the velocity of players are small, and the probability of achieving an all-C state monotonically decreases with the velocity for a fixed payoff parameter.
In our model, however, the maximum cooperation level does not occur at the limit $v\rightarrow 0$, when the movement of players promotes cooperation for modest values of $R$, and a non-monotonic dependence of the cooperator frequency on the velocity of players can be observed in Fig. \ref{fcv:b} and Fig. \ref{fcv:e}. Compared with the result in Ref. \cite{Meloni2009}, this phenomenon can be explained by the difference between the rules of movement.
In Ref. \cite{Meloni2009}, the network of contacts is continuously changing, because individual directions are controlled by N-independent random variables. As a result, randomness among partnerships can be preserved all the time.
But in the present work, a static network of interactions is gradually developed, when individuals successfully align themselves with neighbors. Such fixed partnerships are maintained by the cancellation of periodic boundary conditions, which allows the coexistence of cooperators and defectors.
As shown in Fig. \ref{typicalevo}, the range of alignment interaction fluctuates only when $t<500$, while the variance of neighbors causes a sharp transition of the cooperator frequency.
In the process of direction alignment, the larger the value of $v$, the higher the probability for each individual to encounter different neighbors.
Different with the work in Ref. \cite{Meloni2009}, the value of $v$ also determines how long random partnerships persist. In our work, for a fixed $R$, the larger the value of $v$, the sooner the system is expected to achieve static neighborhoods.
It is not easy to describe how cooperation is promoted by small values of $v$, and a heuristic explanation is that a low degree of migration can trigger the expansion of cooperator clusters, as suggested in Ref. \cite{YangArx}.
In our model, the cooperation level for $v=10^{-6}$ is near to that for $v=0$.
And for large values of $v$, cooperative clusters may be destroyed by the movement of players, making cooperators vulnerable to defectors.
Thus similar to the so called evolutionary coherence resonance \cite{Perc2006a,Perc2006}, the maximum level of cooperation can be induced by an optimal amount of randomness, which is determined by the absolute velocity $v$ of players.
In addition, results in Fig. \ref{fcv} have also shown that the movement of players can inhibit cooperation for a small (large) value of $R$.
Next, we will make discussions about the role of $R$ in the evolution of cooperation.

\begin{figure}[htbp]
\centering
\subfigure[]{\label{fc-R:subfig:c}
\includegraphics[height=5cm,width=5cm]{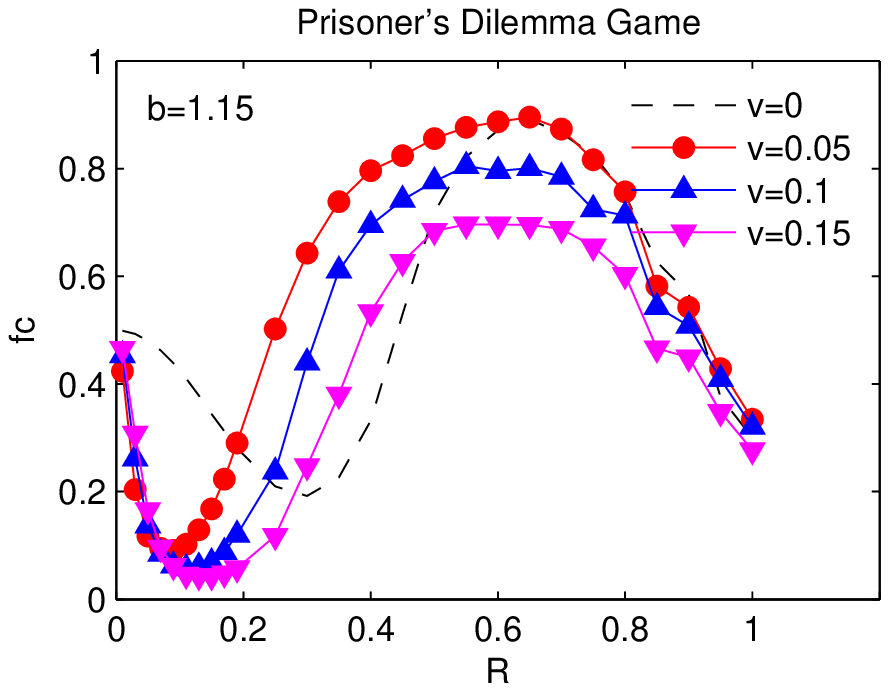}}
\subfigure[]{\label{fc-R:subfig:d}
\includegraphics[height=5cm,width=5cm]{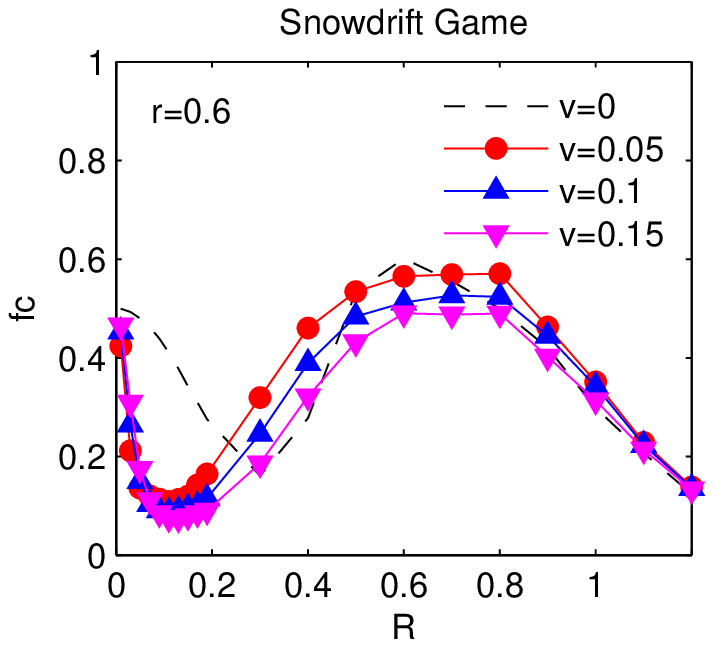}}
\subfigure[]{ \label{fc-R:subfig:a}
\includegraphics[height=5cm,width=5cm]{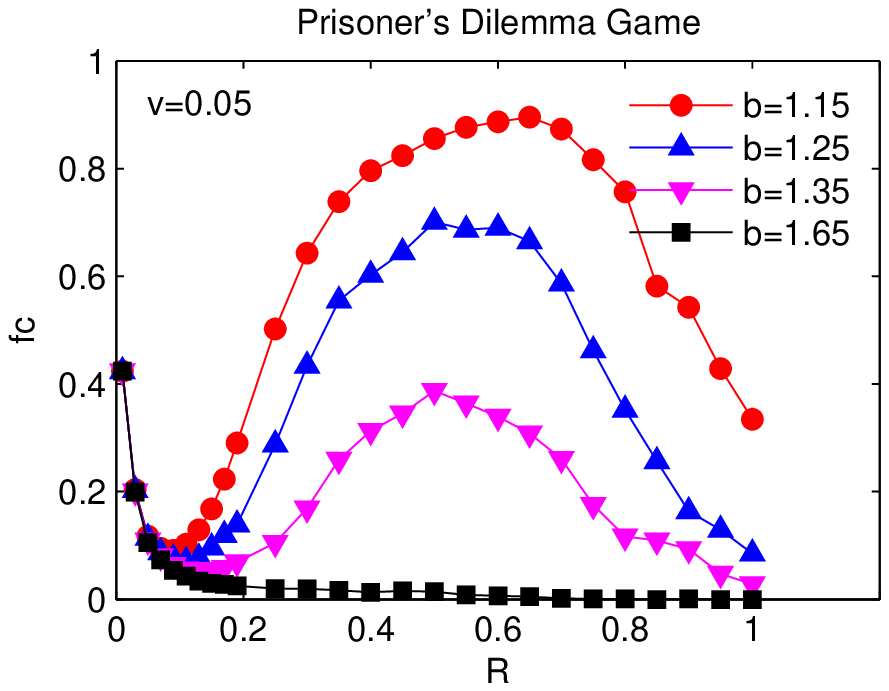}}
\subfigure[]{\label{fc-R:subfig:b}
\includegraphics[height=5cm,width=5cm]{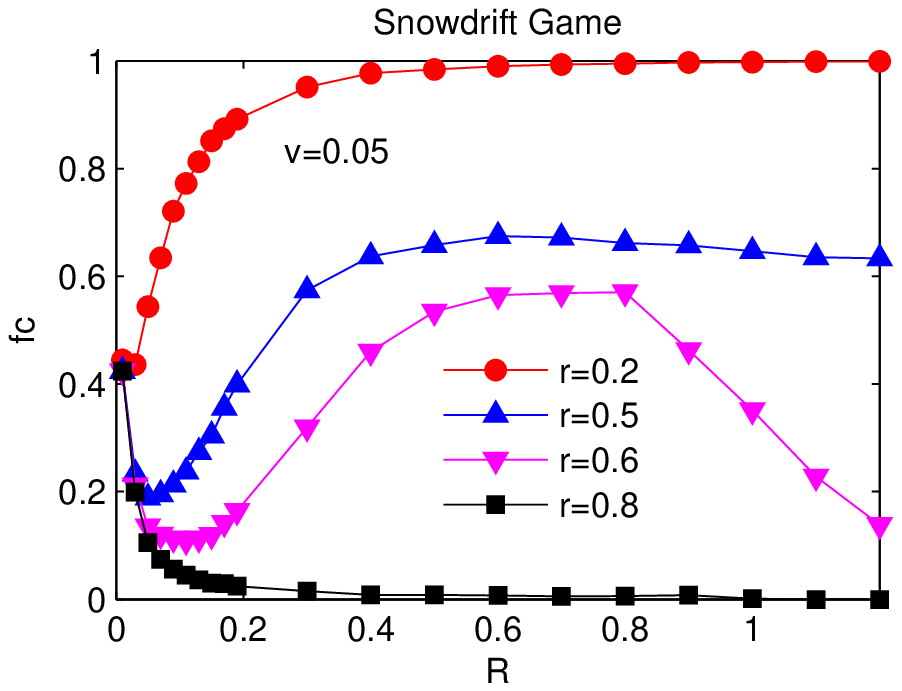}}
\caption{The cooperator frequency $fc$ versus the radius $R$ for different values of $b$ (PD), $r$ (SD) and $v$.}\label{fc-R:subfig}
\end{figure}

In Fig. \ref{fc-R:subfig:c} and Fig. \ref{fc-R:subfig:d},
we show that the cooperator frequency $fc$ varies with the radius $R$ for different values of $v$.
It displays that the proportion of cooperators monotonously decreases with $R$, until the radius exceeds a certain value, and for $R<0.2$, the maximum of $fc$ appears at $v=0$.
Note in the current work, interaction neighborhoods are determined by the radius $R$ at each time step.
For near-zero values of $R$, there are few links among individuals in the instant network.
As pairwise interactions increase with $R$,
it is hard for isolated cooperators to resist the invasion of defectors.
When players are allowed to move, defectors have more chances to exploit cooperators. Then for small $R$, defection becomes dominant in the population, and the movement of players inhibits the evolution of cooperation.
But for larger values of $R$, cooperators are expected to get together, and the introduction of mobility causes a more rapid increase of $fc$. As shown in Fig. \ref{fc-R:subfig:c} and Fig. \ref{fc-R:subfig:d}, the curves of $fc$ for $v>0$ depart around $R=0.09$, and then begin increasing with $R$ in both games.
For each value of $v$, the increase of $R$ induces a resonance-like phenomena, and the cooperator frequency $fc$ reaches a maximum around $R=0.6$. This finding indicates that intermediate values of $R$ are most favorable for cooperation, since the system approaches a fully connected network for extremely large $R$ in the stationary state.
It also helps explain why the movement of players fails to give evident enhancement to the cooperation level at large values of $R$. Indeed, for large $R$, the network of interactions is almost static, since individuals can quickly align themselves with neighbors. As a result, though the maximum of $fc$ decreases with $v$, the curves of $fc$ for different values of $v$ gradually merge at large values of $R$.
In Fig. \ref{fc-R:subfig:a} and Fig. \ref{fc-R:subfig:b}, we show the dependence of the cooperation level on the radius $R$ for different values of $b$ ($r$). It displays that the resonance-like phenomena is greatly influenced by the payoff parameter, and for a fixed $R$, the maximum of $fc$ decreases with $b$ ($r$). For large $b$ ($r$), the cooperator frequency $fc$ monotonously decreases with $R$, and the maximum level of cooperation appears at the limit $R\rightarrow 0$. But for $r=0.2$, the system can reach an absorbing state of all cooperators. This is because the SD is more favorable for cooperators than the PD.

\begin{figure}[htbp]
\centering
\subfigure[]{\label{dens_pd}
\includegraphics[height=5cm, width=5cm]{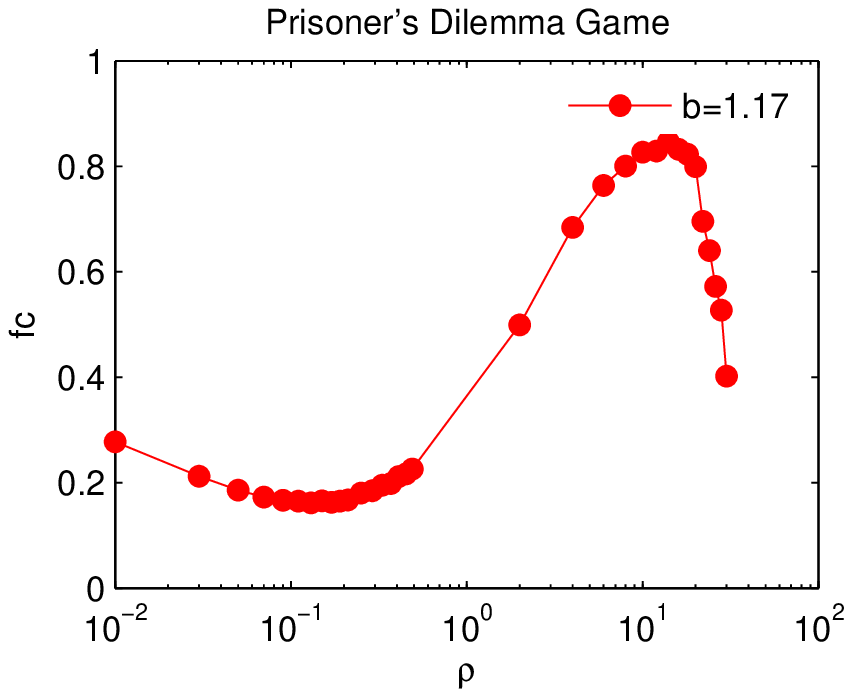}}
\subfigure[]{\label{dens_sd}
\includegraphics[height=5cm, width=5cm]{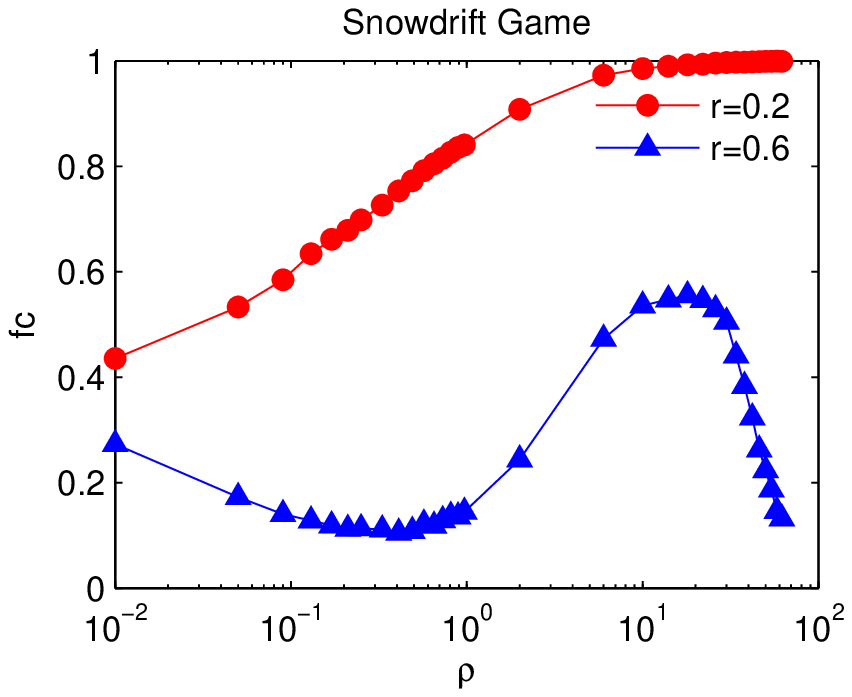}}
\caption{The cooperator frequency $fc$ versus the density $\rho$ for $R=0.5$, $v=0.05$. A logarithmic scale is used for the $X$ axis.} \label{dens}
\end{figure}

In Fig. \ref{dens}, we plot the cooperator frequency $fc$ against the initial density $\rho$ for fixed payoff parameters
when $R=0.5$, $v=0.05$.
One can see that the behavior of $fc$ caused by the variance of $\rho$ is similar to that shown in Fig. \ref{fc-R:subfig}.
This is because both $R$ and $\rho$ can influence the size of neighborhood.
For instance, when $t=0$, the average degree of the interaction network can be written as $<k>=NR^2\pi/L^2=\rho R^2\pi$. When the players are located on the vertices of a fixed network, previous results have shown the resonant behavior of the cooperator frequency around certain values of the average degree \cite{Tang2006}. And our work can be viewed as an extension to the dynamic interaction network that appears during the movement of players.
For small $\rho$, all agents are widely dispersed in the plane,
and cooperators cannot get enough support from cooperative neighbors.
Though the chance of forming cooperative clusters increases with $\rho$, the proportion of cooperators monotonously decreases until $\rho>0.13$, as shown in the inset of Fig. \ref{dens_pd}.
Large values of $\rho$ are also harmful to cooperators. This is because the mean field situation is nicely recovered for large $\rho$, in which interactions almost take place among each pair of players.
Between these two limits, one can find that the cooperation level can reach the maximum for moderate values of $\rho$.
It has been reported that the cooperation level can reach a peak at some values of $\rho$, when the players are running in a square with periodic boundary conditions \cite{Meloni2009}. And our results indicate the importance of the initial density $\rho$ to the evolution of cooperation, even when the boundary restrictions are removed.
In the SD, one can find the similar phenomenon that observed in the PD: when $r=0.6$, the cooperator frequency $fc$ first decreases for small $\rho$, then increases with $\rho$ until reaching the maximum, and decreases for large $\rho$.
Note when $r=0.2$, the cooperator frequency $fc$ monotonously increases with $\rho$, and the system can reach an absorbing state of full cooperators at last.

\section{Conclusion}
To summarize,
we have investigated the PD game and the SD game in a population of mobile players, which move at a constant speed $v$ while interacting with neighbors within a fixed radius $R$.
Through alignment of traveling direction, individuals develop static interaction neighborhoods during movement.
Numerical simulations show that cooperation can be maintained with simple strategies, provided the payoff parameter and the velocity are small. Compared with the case for $v=0$, cooperation can be enhanced by the movement of players, and there is an optimal value of $v$ to induce the maximum cooperation level. However, such enhancement of cooperation can only be observed for modest values of $R$. This is because cooperators have less chances to form clusters for small $R$, while interaction neighborhoods approach a static, full connected network for large $R$. Besides, when $v$ is fixed, our results suggest that intermediate values of $R$ and initial population densities $\rho$ are most favorable for cooperation, since both $R$ and $\rho$ can influence the size of neighborhood.
The resonance-like phenomenon caused by $R$ and $\rho$ is also affected by the payoff parameter,
and for a small cost-to-benefit ratio, the system can reach an absorbing state of full cooperators.
Our results may be relevant for understanding the role of information flows in cooperative, multi-vehicle systems \cite{Olfati-Saber2007}.

The work is jointly supported by the National Natural Science Foundation of China (61004088), the Key Foundation for Basic Research from Science and Technology Commission of Shanghai (09JC1408000) and the Aeronautic Science Foundation of China(20100157001).


\begin{thebibliography}{00}
\bibitem{Okasha}
S.~Okasha, Biological Altruism, The Stanford Encyclopedia of Philosophy, http://plato.stanford.edu/archives/win2009/entries/altruism-biological/.

\bibitem{Doebeli2005}
M.~Doebeli, C.~Hauert, Models of cooperation based on the prisoner's dilemma
  and the snowdrift game, Ecol. Lett. 8~(7) (2005) 748--766.

\bibitem{Nowak2006}
M.~A. Nowak, Five rules for the evolution of cooperation, Science 314~(5805)
  (2006) 1560--1563.

\bibitem{Nowak1993}
M.~A. Nowak, R.~M. May, The spatial dilemmas of evolution, Int. J. Bifurcation
  Chaos 3 (1993) 35--78.

\bibitem{Szabo1998}
G.~Szab{\'o}, C.~T{\H o}ke, Evolutionary prisoner's dilemma game on a square
  lattice, Phys. Rev. E 58 (1998) 69--73.

\bibitem{Szolnoki2009a}
A.~Szolnoki, J.~Vukov, G.~Szab{\'o}, Selection of noise level in strategy
  adoption for spatial social dilemmas, Phys. Rev. E 80~(5) (2009) 056112.

\bibitem{Szabo2009}
G.~Szab{\'o}, A.~Szolnoki, J.~Vukov, Selection of dynamical rules in spatial
  prisoner's dilemma games, Epl 87~(1) (2009) 18007.

\bibitem{Santos2008}
F.~C. Santos, M.~D. Santos, J.~M. Pacheco, Social diversity promotes the
  emergence of cooperation in public goods games, Nature 454~(7201) (2008)
  213--217.

\bibitem{Yang2009}
H.~X. Yang, W.~X. Wang, Z.~X. Wu, Y.~C. Lai, B.~H. Wang, Diversity-optimized
  cooperation on complex networks, Phys. Rev. E 79~(5) (2009) 056107.

\bibitem{Wu2009}
Z.~X. Wu, Z.~H. Rong, P.~Holme, Diversity of reproduction time scale promotes
  cooperation in spatial prisoner's dilemma games, Phys. Rev. E 80~(3) (2009)
  036106.

\bibitem{Chen2009}
Y.~Z. Chen, Z.~G. Huang, S.~J. Wang, Y.~Zhang, Y.~H. Wang, Diversity of
  rationality affects the evolution of cooperation, Phys. Rev. E 79~(5) (2009)
  055101.

\bibitem{Perc2008}
M.~Perc, A.~Szolnoki, Social diversity and promotion of cooperation in the
  spatial prisoner's dilemma game, Phys. Rev. E 77 (2008) 011904.

\bibitem{VanSegbroeck2009}
S.~Van~Segbroeck, F.~C. Santos, T.~Lenaerts, J.~M. Pacheco, Reacting
  differently to adverse ties promotes cooperation in social networks, Phys.
  Rev. Lett. 102~(5) (2009) 058105.

\bibitem{Santos2005}
F.~C. Santos, J.~M. Pacheco, Scale-free networks provide a unifying framework
  for the emergence of cooperation, Phys. Rev. Lett. 95 (2005) 098104.

\bibitem{Gomez-Gardenes2007}
J.~G{\'o}mez-Garde\~{n}es, M.~Campillo, L.~M. Flor{\'i}a, Y.~Moreno, Dynamical
  organization of cooperation in complex topologies, Phys. Rev. Lett. 98 (2007)
  108103.

\bibitem{Ren2007}
J.~Ren, W.~X. Wang, F.~Qi, Randomness enhances cooperation: A resonance-type
  phenomenon in evolutionary games, Phys. Rev. E 75 (2007) 045101.

\bibitem{Chen2008}
X.~J. Chen, L.~Wang, Promotion of cooperation induced by appropriate payoff
  aspirations in a small-world networked game, Phys. Rev. E 77 (2008) 017103.

\bibitem{Du2009}
W.~B. Du, X.~B. Cao, M.~B. Hu, H.~X. Yang, H.~Zhou, Effects of expectation and
  noise on evolutionary games, Physica A 388 (2009) 2215--2220.

\bibitem{Pena2009}
J.~Pe\~{n}a, H.~Volken, E.~Pestelacci, M.~Tomassini, Conformity hinders the
  evolution of cooperation on scale-free networks, Phys. Rev. E 80~(1) (2009)
  016110.

\bibitem{Szolnoki2008}
A.~Szolnoki, M.~Perc, Coevolution of teaching activity promotes cooperation,
  New J. Phys. 10 (2008) 043036.

\bibitem{Szolnoki2009}
A.~Szolnoki, M.~Perc, Promoting cooperation in social dilemmas via simple
  coevolutionary rules, Eur. Phys. J. B 67~(3) (2009) 337--344.

\bibitem{Moyano2009}
L.~G. Moyano, A.~S{\'a}nchez, Evolving learning rules and emergence of
  cooperation in spatial prisoner's dilemma, J.Theor.Biol. 259 (2009) 84--95.

\bibitem{Cardillo2010}
A.~Cardillo, J.~G{\'o}mez-Garde{\~n}es, D.~Vilone, A.~S{\'a}nchez, Co-evolution
  of strategies and update rules in the prisoner's dilemma game on complex
  networks, New J. Phys. 12 (2010) 103034.

\bibitem{Zimmermann2004}
M.~G. Zimmermann, V.~M. Egu{\'\i}luz, M.~San~Miguel, Coevolution of dynamical
  states and interactions in dynamic networks, Phys. Rev. E 69 (2004) 065102.

\bibitem{Santos2006}
F.~C. Santos, J.~M. Pacheco, T.~Lenaerts, Cooperation prevails when individuals
  adjust their social ties, PLOS Comp. Biol. 2 (2006) 1284--1291.

\bibitem{Szolnoki2008a}
A.~Szolnoki, M.~Perc, Z.~Danku, Making new connections towards cooperation in
  the prisoner's dilemma game, EPL 84~(5) (2008) 50007.

\bibitem{Chen2009a}
X.~J. Chen, F.~Fu, L.~Wang, Social tolerance allows cooperation to prevail in
  an adaptive environment, Phys. Rev. E 80~(5) (2009) 051104.

\bibitem{Fu2008}
F.~Fu, C.~Hauert, M.~A. Nowak, L.~Wang, Reputation-based partner choice
  promotes cooperation in social networks, Phys. Rev. E 78 (2008) 026117.

\bibitem{Poncela2008}
J.~Poncela, J.~G{\'o}mez-Garde\~{n}es, L.~M. Flor{\'i}a, A.~S{\'a}nchez,
  Y.~Moreno, Complex cooperative networks from evolutionary preferential
  attachment, Plos One 3~(6) (2008) e2449.

\bibitem{Poncela2009}
J.~Poncela, J.~G{\'o}mez-Garde\~{n}es, A.~Traulsen, Y.~Moreno, Evolutionary
  game dynamics in a growing structured population, New J. Phys. 11 (2009)
  083031.

\bibitem{Szolnoki2009c}
A.~Szolnoki, M.~Perc, Emergence of multilevel selection in the prisoner's
  dilemma game on coevolving random networks, New J. Phys. 11 (2009) 093033.

\bibitem{Szolnoki2009b}
A.~Szolnoki, M.~Perc, Resolving social dilemmas on evolving random networks,
  Epl 86~(3) (2009) 30007.

\bibitem{Szabo2007}
G.~Szab{\'o}, G.~F{\'a}th, Evolutionary games on graphs, Phys. Rep. 446 (2007)
  97--216.

\bibitem{Perc2010}
M.~Perc, A.~Szolnoki, Coevolutionary games-a mini review, Biosystems 99~(2)
  (2010) 109--125.

\bibitem{Kleinberg2007}
J.~Kleinberg, Computing: The wireless epidemic, Nature 449~(7160) (2007)
  287--288.

\bibitem{Brockmann2006}
D.~Brockmann, L.~Hufnagel, T.~Geisel, The scaling laws of human travel, Nature
  439 (2006) 462--465.

\bibitem{Gonzalez2008}
M.~C. Gonz{\'a}lez, C.~A. Hidalgo, A.~L. Barab{\'a}si, Understanding individual
  human mobility patterns, Nature 453 (2008) 779--782.

\bibitem{Song2010}
C.~M. Song, Z.~H. Qu, N.~Blumm, A.~L. Barabasi, Limits of predictability in
  human mobility, Science 327~(5968) (2010) 1018--1021.

\bibitem{Gonzalez2006}
M.~C. Gonz{\'a}lez, P.~G. Lind, H.~J. Herrmann, System of mobile agents to
  model social networks, Phys. Rev. Lett. 96 (2006) 088702.

\bibitem{Vainstein2007}
M.~H. Vainstein, A.~T.~C. Silva, J.~J. Arenzon, Does mobility decrease
  cooperation?, J. Theor. Biol. 244 (2007) 722--728.

\bibitem{Guan2007}
J.~Y. Guan, Z.~X. Wu, Y.~H. Wang, Evolutionary snowdrift game with disordered
  environments in mobile societies, Chinese Phys 16~(12) (2007) 3566--3570.

\bibitem{Sicardi2009}
E.~A. Sicardi, H.~Fort, M.~H. Vainstein, J.~J. Arenzon, Random mobility and
  spatial structure often enhance cooperation, J. Theor. Biol. 256~(2) (2009)
  240--246.

\bibitem{Hauert2004}
C.~Hauert, M.~Doebeli, Spatial structure often inhibits the evolution of
  cooperation in the snowdrift game, Nature 428 (2004) 643--646.

\bibitem{Aktipis2004}
C.~A. Aktipis, Know when to walk away: contingent movement and the evolution of
  cooperation, J. Theor. Biol. 231 (2004) 249--260.

\bibitem{Helbing2009}
D.~Helbing, W.~J. Yu, The outbreak of cooperation among success-driven
  individuals under noisy conditions, Proc. Natl. Acad. Sci. USA 106~(10)
  (2009) 3680--3685.

\bibitem{Jiang2010}
L.~L. Jiang, W.~X. Wang, Y.~C. Lai, B.~H. Wang, Role of adaptive migration in
  promoting cooperation in spatial games, Phys. Rev. E 81~(3) (2010) 036108.

\bibitem{Vicsek1995}
T.~Vicsek, A.~Czir{\'o}k, E.~Ben-Jacob, I.~Cohen, O.~Shochet, Novel type of
  phase transition in a system of self-driven particles, Phys. Rev. Lett. 75
  (1995) 1226--1229.

\bibitem{Nagy2007}
M.~Nagy, I.~Daruka, T.~Vicsek, New aspects of the continuous phase transition
  in the scalar noise model (snm) of collective motion, Physica A 373 (2007)
  445--454.

\bibitem{Dossetti2009}
V.~Dossetti, F.~J. Sevilla, V.~M. Kenkre, Phase transitions induced by complex
  nonlinear noise in a system of self-propelled agents, Phys. Rev. E 79 (2009)
  051115.

\bibitem{Chen2011}
Z.~Chen, J.~X. Gao, Y.~Z. Cai, X.~M. Xu, Evolutionary prisoner’s dilemma game
  in flocks, Physica A 390 (2011) 50--56.

\bibitem{Dai2007}
X.~B. Dai, Z.~Y. Huang, C.~X. Wu, Evolution of cooperation among interacting
  individuals through molecular dynamics simulations, Physica A 383 (2007)
  624--630.

\bibitem{Meloni2009}
S.~Meloni, A.~Buscarino, L.~Fortuna, M.~Frasca, J.~G{\'o}mez-Garde{\~n}es, V.~Latora,
  Y.~Moreno, Effects of mobility in a population of prisoner's dilemma players,
  Phys. Rev. E 79 (2009) 067101.

\bibitem{YangArx}
H.~X. Yang, W.~X. Wang, B.~H. Wang, Universal role of migration in the evolution of cooperation, arXiv:1005.5453v1 [physics.soc-ph]

\bibitem{Perc2006a}
M.~Perc, M.~Marhl, Evolutionary and dynamical coherence resonances in the pair
  approximated prisoner's dilemma game, New J. Phys. 8 (2006) 142.

\bibitem{Perc2006}
M.~Perc, Coherence resonance in a spatial prisoner's dilemma game, New J. Phys.
  8 (2006) 22.

\bibitem{Tang2006}
C.~L. Tang, W.~X. Wang, X.~Wu, B.~H. Wang, Effects of average degree on
  cooperation in networked evolutionary game, Eur. Phys. J. B 53~(3) (2006)
  411--415.

\bibitem{Olfati-Saber2007}
R.~Olfati-Saber, J.~A. Fax, R.~M. Murray, Consensus and cooperation in
  networked multi-agent systems, Proc. IEEE 95 (2007) 215--233.
\end{thebibliography}
\end{document}